\documentclass[prb,aps,twocolumn,showpacs,showkeys,reprint,superscriptaddress,amsmath,amssymb]{revtex4}
\usepackage{graphicx}
\usepackage{color}
\usepackage{amssymb,amsmath}
\usepackage{comment}

\usepackage{verbatim}

\usepackage{hyperref}
\hypersetup{
bookmarks=false,
pdftitle={Microwave photonics}
anchorcolor=blue,
colorlinks=true,
citecolor=blue,
urlcolor=blue,
linkcolor=blue}
\usepackage{marginnote}
\usepackage[author=Juanjo]{pdfcomment}
\usepackage{xcolor}
\newcommand{\mycomment}[1]{}

\begin{document}

\title{Microwave photonics with Josephson junction arrays: \\ 
negative refraction index and entanglement through disorder}

\author{David Zueco} 

\affiliation{Instituto de Ciencia de Materiales de Arag\'on y Departamento de F\'{\i}sica de la Materia Condensada, CSIC-Universidad de Zaragoza, E-50009 Zaragoza, Spain.}
\affiliation{Fundaci\'on ARAID, Paseo Mar\'{\i}a Agust\'{\i}n 36, 50004 Zaragoza, Spain}

\author{Juan Jos\'e Mazo}

\affiliation{Instituto de Ciencia de Materiales de Arag\'on y Departamento de F\'{\i}sica de la Materia Condensada, CSIC-Universidad de Zaragoza, E-50012 Zaragoza, Spain.}

\author{Enrique Solano} 

\affiliation{Departamento de Qu\'{\i}mica F\'{\i}sica, Universidad del
  Pa\'{\i}s Vasco UPV/EHU, Apdo. 644, 48080 Bilbao, Spain}

\affiliation{IKERBASQUE, Basque Foundation for Science, Alameda Urquijo 36, 48011 Bilbao, Spain}

\author{Juan Jos\'e Garc\'{\i}a-Ripoll}

\affiliation{Instituto de F\'{\i}sica Fundamental, IFF-CSIC, Serrano 113-bis, 28006 Madrid, Spain}

\date{\today}

\begin{abstract}
  We study different architectures for a photonic crystal in the
  microwave regime based on superconducting transmission lines
  interrupted by Josephson junctions, both in one and two
  dimensions. A study of the scattering properties of a single
  junction in the line shows that the junction behaves as a perfect
  mirror when the photon frequency matches the Josephson plasma
  frequency. We generalize our calculations to periodic arrangements
  of junctions, demonstrating that they can be used for tunable band
  engineering, forming what we call a \textit{quantum circuit
    crystal}. Two  applications are discussed in detail.  In a
  two-dimensional structure we demonstrate the phenomenon of
  negative refraction.  We  finish by studying  the creation of stationary entanglement between two superconducting qubits interacting through a disordered media.
\end{abstract}

\maketitle
                             

\section{Introduction}

Circuit QED~\cite{You2011} is quantum optics on a superconducting chip: a solid state analogue of cavity QED in which superconducting resonators and qubits act as optical cavities and artificial atoms. After successfully reproducing many key experiments from the visible regime ---qubit-photon strong coupling and Rabi oscillations~\cite{Wallraff2004}, Wigner function reconstruction~\cite{Hofheinz2009}, cavity-mediated qubit-qubit coupling~\cite{Majer2007}, quantum algorithms~\cite{Dicarlo2009} or Bell inequalities measurement~\cite{Ansmann2009}---, and improving the quality factors of qubits and cavities, c-QED establishes as an alternative to standard quantum optical setups.

The next challenge in the field is the development of \textit{quantum
  microwave photonics} in the Gigahertz regime. The scope is the
generation, control and measurement of propagating photons,
contemplating all its possibilities as carriers of quantum information
and mediators of long distance correlations. The natural framework is
that of active and passive \textit{quantum metamaterials}, with open
transmission lines to support propagation of photons and embedded
circuits to control them \cite {Shen2007a, Rakhmanov2008,
  Zagoskin2009, Liao2010, Hutter2011}.  Qubits can be a possible ingredient in
these metamaterials. A two-level system may act as a saturable mirror
for resonant çphotons~\cite{Shen2005, Zhou2008a, Zhou2008b}, as it has been demonstrated in
a breakthrough experiment with flux qubits~\cite{Astafiev2010a},
continued by further demonstrations of single photon
transistors~\cite{hoi2011}, and electromagnetically induced
transparency~\cite{Astafiev2010}. These groundbreaking developments,
together with theoretical studies of band engineering using
qubits~\cite{Shen2007a, Rakhmanov2008, Zagoskin2009} and Josephson
junction arrays~\cite{Hutter2011}, and recent developments in the
field of photodetection~\cite{Peropadre11,Romero2009, Chen2011}, provide solid foundations for this rapidly growing field. It is important to contrast these developments with alternative setups in the high-energy microwave regime (Terahertz)~\cite{Savelev2005, Savelev2006, Savelev2010}, which differ both in the architecture and the scope.

In this work, we advocate an alternative architecture for both passive and active quantum metamaterials based on transmission lines with embedded Josephson junctions (JJ). Adopting a bottom-up approach, we first study the scattering of travelling photons through a single junction, the simplest and most fundamental element in superconducting technologies. It is shown that, in the few photon limit, the linearized junction acts as a perfect mirror for resonant photon. Starting from the single JJ scattering matrix, we show how to engineer metamaterials using periodic arrangements of junctions both in one and two-dimensional transmission line networks. Compared to previous approaches, this combines the travelling nature and flexible geometry of photons in transmission lines~\cite{Shen2005}, and instead of qubits~\cite{Shen2005,Rakhmanov2008,Zagoskin2009} it relies on the simple and robust dynamics of a linearized junction~\cite{Hutter2011}. Previous proposals lacked one of these two ingredients.

The simplicity of this setup opens the door to multiple short-term applications. In this manuscript we discuss mainly two. The first one is the observation of a negative index of refraction in a two-dimensional circuit crystal. This would be achieved by injecting an appropriate microwave in a square network of transmission lines, where only half of it is populated with embedded junctions. Second and most important, we study the interaction between qubits in a disordered quantum metamaterial, showing that a sufficiently large disorder can support the generation of entanglement between two distant flux qubits.  The main conclusion of this study is that different topics in the fields of metamaterials and localization, usually discussed in the classical or {\it many} photon level can be realized in the few photon limit inside the field of circuit QED.

The paper is structured as follows. In the Sec.~\ref{sec:scatterer} we
discuss the scattering through a single JJ in the linear regime,
computing its reflection and transmission coefficients. Using these
results, Sec.~\ref{sec:qcc} develops the theory of transmission lines
with periodically embedded Josephson junctions. We show how to compute
and engineer the band structure of these photonic crystals and, as
application we discuss the implementation of a negative index of
refraction in two-dimensional arrangements. In Sec.~\ref{sec:qme} we
study the coupling between qubits and those structured lines. We develop an analytical theory that models the interaction and dissipation of superconducting qubits in a network of JJ and transmission lies, within the master equation formalism.  This theory is then applied to the study of the steady entanglement between two separated qubits that interact with a structured line where it has been induced disorder.  We finish with the conclusions, while some technical aspects are elaborated in the appendices.


\section{Josephson junction as a scatterer}
\label{sec:scatterer}

JJs are the most versatile nonlinear element in circuit QED. Either
alone, or in connection with extra capacitors or junctions, they form
all types of superconducting qubits to
date~\cite{Makhlin2001}. Moreover, in recent years they have also been
used inside cavities to shape and control confined photons,
dynamically tuning the mode structure~\cite{Castellanos-Beltran2008,
  Wilson2011}, enhancing the light-matter
coupling~\cite{Niemczyk2010,Forn-Diaz2010}, or exploiting their
nonlinearity in resonators~\cite{Ong2010}.  Junctions have also been
sugested as control elements for propagating photons in two different
ways. One approach consist of SQUIDs or charge qubit arrays  to control the photon dispersion relation forming one dimensional  {\it quantum metamaterials} \cite{Rakhmanov2008,  Zagoskin2009, Nation2009, Hutter2011}. The other alternative  relies on the single photon scattering  by superconducting qubits~\cite{Astafiev2010a, Shen2005}, using the fact that two level systems act as perfect mirror whenever the incident photon frequency and the qubit splitting equals.

In the following we combine these ideas, providing both a uniform theoretical framework to study the interaction of Josephson junctions with propagating photons and a scalable architecture to construct \textit{quantum metamaterials} by periodic arrangements of these junctions. Just like in the case of qubits, we expect that a JJ in an open line may act as a perfect scatterer of propagating photons, where now the resonant frequency is given by the JJ plasma frequency.

\begin{figure}
  \includegraphics[width=\linewidth]{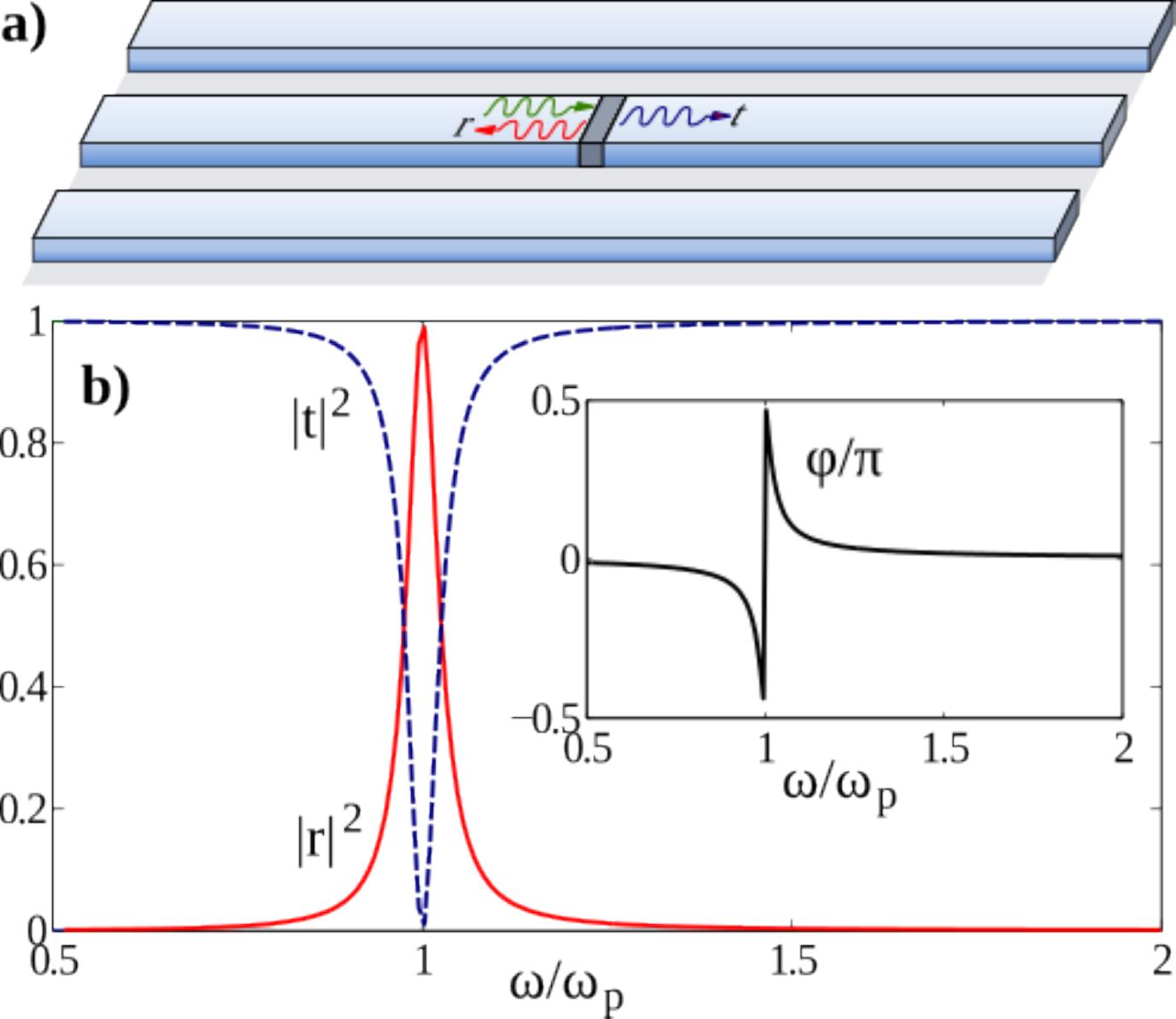}
  \caption{(color online) (a) An open transmission line interrupted by a Josephson junction. (b) Reflection, $r$, transmission, $t$, and phase of the transmitted beam, $\varphi = \arg t$, vs. incoming photon frequency, in units of the plasma frequency $\omega_p$. We use $Z_0/Z_J=10$.}
\label{fig:jj-sc}
\end{figure}

In our study we will adopt a bottom-up approach starting from the scattering problem of a single junction~[Fig.~\ref{fig:jj-sc}] that interacts with incoming and outgoing microwave packets. The Lagragian for this system combines the one-dimensional field theory for a transmission line with the capacitively-shunted-junction model for the junction~\cite{Niemczyk2010, Ong2010, Bourassa2009}
\begin{align}
\label{tl-jj}
\mathcal L
&=
\frac{1}{2}
\int_{-\infty}^{0_-} {\rm d}x 
\left[
c_0 (\partial_t \phi) ^2 -
\frac{1}{l_0} (\partial_x \phi) ^2\right]
\\ \nonumber
&+
\frac{1}{2}  \left (\frac{\Phi_0}{2 \pi} \right )^2 C_J \left (\frac{d \varphi}{dt} \right) ^2
-\left (\frac{\Phi_0}{2 \pi} \right )
I_C \cos{\varphi}
\\ \nonumber
&+
\frac{1}{2}
\int_{0_+}^{\infty} {\rm d}x 
\left[
c_0 (\partial_t \phi) ^2 -
\frac{1}{l_0} (\partial_x \phi) ^2\right].
\end{align}
The field $\phi(x,t)$ represents flux on the line. The line capacitance and inductance per unit length, $c_0$ and $l_0$, are assumed uniform for simplicity (See Ref.~\onlinecite{Bourassa2009} for generalizations).  The junction, placed at $x=0$, is characterized by a capacitance $C_J$ and a critical current $I_C$ together with the gauge invariant phase, $\varphi$
\begin{equation}
\label{varphi}
\varphi = \Delta \theta - \frac{2 \pi}{\Phi_0} \int_{0_-}^{0_+} {\bf A} ({\bf r},t) \cdot {\rm d} {\bf l} \; ,
\end{equation}
where $\Delta \theta$ is the superconducting phase difference and ${\bf A}({\bf r},t)$ is the vector potential.

It is convenient to introduce the field $\tilde \phi (x,t)$ as the variations over the static flux $\phi^{(0)} (x)$,
\begin{equation}
 \phi (x,t) = \phi^{(0)} (x) + \tilde \phi (x,t),
\end{equation}
and a flux variable $\delta \phi (t)$ associated to the time fluctuations for the flux
across the junction $\delta \phi(t) := \tilde \phi (0_+, t) - \tilde \phi (0_-, t)$ defined as
\begin{equation}
\label{wtvp}
\varphi (t) = \varphi^{(0)}+\frac{2\pi}{\Phi_0} \delta \phi (t)
\end{equation}
Here $\varphi^{(0)}$ stands for the equilibrium solution for the phase and
$V = (\Phi_0/2\pi) \dot \varphi = \dot{\delta \phi}$  is the expected voltage-flux relation~\cite{Devoret1995}.

The fields to the left and to the right of the junction are matched using current conservation, which states that $I_{\rm left}$, $I_{\rm junction}$ and $I_{\rm right}$ are equal at $x=0$,
\begin{equation}
\frac{1}{l_0} \partial_x \phi(0_\pm, t) =
\frac{\Phi_0}{2 \pi} C_J \ddot {\varphi} + I_C \sin{\varphi}
\label{wtcc}
\end{equation}
These two equations may be formally solved, but the result is a
complicated nonlinear scattering problem. In order to get some
analytical understanding of the junction as a scatterer, and since we
are mostly interested in the few photon  regime, we will linearize
equations (\ref{wtcc})  assuming small fluctuations in the junction phase
$\delta \phi$, $\sin \varphi \cong \sin \varphi^{(0)} + \frac{2
  \pi}{\Phi_0}  \cos (\varphi^{(0)}) \; \delta \varphi$, yielding
\begin{equation}
\label{scjjlf}
\frac{1}{l_0} \partial_x \tilde \phi(0_\pm, t) = 
C_J  \, \ddot {\delta  \phi} + \frac{1}{L_J} \, \delta \phi 
\end{equation}
with $L_J=\Phi_0/(2\pi I_C \cos{(\varphi^{(0)})})$.
Besides the static fields are given by $1/l_0 \partial_x \phi^{(0)} (x) = I_c \sin(\varphi^{(0)})$.

In the linearized theory, the stationary scattering solutions can be
written as a combination of incident, reflected and transmitted plane waves:
\begin{equation}
\label{ansatz_jj}
\tilde \phi(x,t) =
A_\phi
\left \{
\begin{array} {cc}
{\rm e}^{i ( k x - \omega t)} + r {\rm e}^{-i ( k x + \omega t)} & (x
< 0)
\\
t {\rm e}^{i ( k x - \omega t)} & (x > 0)
\end{array}
\right .
\end{equation}
where $A_\phi$ is some arbitrary field amplitude, and $r$ and $t$ are the reflexion and transmission coefficients, respectively.  We further assume the scattered waves follow a linear dispersion relation, $\omega=v k$, which is the same outside the junction. Building on the ansatz (\ref{ansatz_jj}) the coefficients are computed yielding,
\begin{equation}
\label{r-jj}
r
=
\frac{1}{1 + i 2 \frac{Z_0}{Z_J} \frac{1}{\bar \omega}
\left ( 
\bar \omega ^2
-1
\right )}
,\quad
t = 1 -r,
\end{equation}
with the rescaled photon frequency, $\bar \omega = \omega / \omega_p$,
with $\omega_p = 1/\sqrt{L_J C_J}$ 
and the impedances of the line and the junction, $Z_J=\sqrt{L_J/C_J}$
and $Z_0=\sqrt{l_0/c_0}$. This formula, which is analogous to the one
for a qubit~\cite{Shen2005,Astafiev2010a}, exhibits perfect reflection
when the photon is on resonance with the junction, $\omega =
\omega_p$, accompanied by the usual phase jump across it
(cf. Fig.~\ref{fig:jj-sc}).

\section{Quantum circuit crystals}
\label{sec:qcc}

We can scale up the previous results, studying periodic arrangements
of junctions both in one and two dimensions. These and other setups
\cite {Shen2007a, Rakhmanov2008,
  Zagoskin2009, Hutter2011} can be seen as a generalization of photonic crystals to the quantum microwave regime, with similar capabilities for controlling
the propagation of photons: engineered dispersion relations, gaps of
forbidden frequencies, localized modes, adjustable group velocities \cite{Nation2009} and index of refraction, and control of the emission and absorption of embedded artificial atoms (i.e. improved cavities)~\cite{Joannopoulos}.

In the following,  we will use the linearized scattering
theory discussed so far.
Regarding possible nonlinear corrections, we expand
the cosine term in (\ref{tl-jj}),  $1/(2 L_J) \delta \phi^2
  (1 - (2 \pi)^2/(12 /\Phi_0^2) \delta \phi^2)$, containing both the
  linear contribution and the first nonlinear correction. Since we are working in the
  single-photon regime, we can replace $\delta\phi^2$ by its
  fluctuations on the vacuum which are proportional to the
  discontinuity of the wavefunction on that point $\delta \phi \sim
  \sqrt{\hbar Z_0 /2}$. All together implies a correction of $0.2 \%$
  compared to the linear contribution.   If we simply view these
  nonlinear corrections as an inductance dispersion we conclude that 
we can safely neglect them, since as we will show in section \ref{sec:ed} such a
dispersion hardly affects the transport properties.

\subsection {One dimensional circuit crystals}

The simplest possible instance of a quantum circuit crystal consists
of a unit cell with $N$ junctions that repeat periodically in a one
dimensional line.  The Lagrangian is a generalization of
Eq.~(\ref{tl-jj}), combining the junctions together with the
intermediate line fields.  In the 1D case there are no additional
constraints on the flux and at equilibrium
$\varphi_j^{(0)}= \phi ^{(0)} (x) = 0$ minimizes the energy [see below
Eq. (\ref{fqxy})]. The scattering problem is translationally invariant
and its eigensolutions are determined by the transfer matrix of the
unit cell, $T_{\mathrm{cell}},$ which relates the field at both sides,
$\tilde \phi_{L,R}(x)=a_{L,R}e^{ikx}+b_{L,R}e^{-ikx},$ through
\begin{equation}
\label{aTa}
  \left(
    \begin{array}{l}
      a_R \\ b_R
    \end{array}
  \right) =
  T_{\mathrm{cell}}(\omega) \left(
    \begin{array}{l}
      a_L \\ b_L
    \end{array}
  \right).
\end{equation}
For a setup with junctions and free lines, the transfer matrix has the form $T_{\mathrm{cell}} = \prod_{i=1}^N T_i D_i$, where $T_i$ is the transfer matrix of the $i$-th junction and $D_i$ is the free propagator through a distance $d_i$ \cite{Shen2007a}
\begin{equation}
\label{Tmat}
T_i = \left (\begin{array}{cc} 1/t_i^* & - r_i^*/t_i^* \\ -r_i/t_i & 1/t_i \end{array} \right ),\;
  D_i = \left (\begin{array}{cc} e^{i\omega d_{i}/v} & 0 \\ 0 & e^{-i\omega d_{i}/v} \end{array} \right ).
\end{equation}
The stationary states are given by Bloch waves, which are eigenstates
of the displacement operator between equivalent sites. Since this
operator is unitary, the eigenvalue can only be a phase,
$\phi(x_{j+1}) = \exp(i p) \phi(x_j)$, which we associate with the
quasimomentum $p = k d$ with $d$ the intercell distance. Moreover, as any two equivalent points in the lattice are related by the transfer matrix and some free propagators, the result is an homogeneous system of linear equations whose solution is found by imposing  $\mathrm{det}[T_{\mathrm{cell}}(\omega) - e^{i p}]=0$ or,
\begin{equation}
\label{bandscond1D}
2 \cos (p) = {\rm Tr} [\hat T_{\rm cell}(\omega)].
\end{equation}

\begin{figure}
  \includegraphics[width=\linewidth]{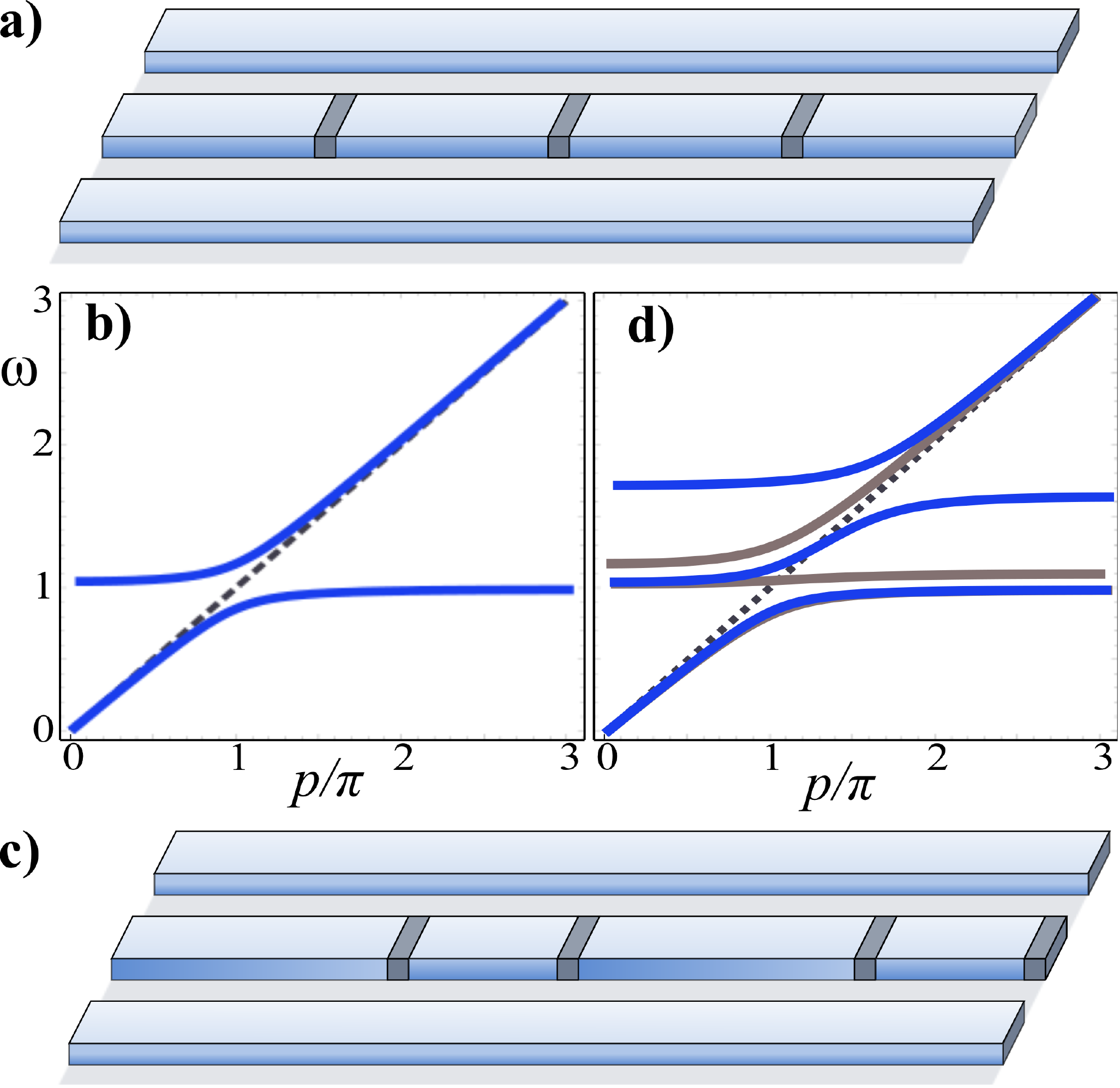}
  \caption{(color online) Photonic crystals with one (a) or two (c) junctions per
    unit cell and their respective energy bands (b and d)
    vs. quasimomentum $p$. We use $Z_0/Z_J = 10$ and $d=0.1 \lambda_J$
    with $\lambda_J$ the typical wave lenght
    ($\lambda = 2 \pi v /\omega_J$. In (d) $\omega_p'/\omega_p=0.6$ (blue) or $0.9$ (gray) and and the distance inside the unit cell is $0.01/d$. Notice in (d) that in the lower band the gray and blue line are indistinguishable.}
  \label{fig:bands}
\end{figure}

As an example, Fig.~\ref{fig:bands} shows the dispersion relation
$\omega(p)$ for  two simple arrangements. The first one is a line with
identical Josephson frequency $\omega_p$ and impedance $Z_J$, evenly
spaced a distance $d$~[Fig.~\ref{fig:bands}a]. The second one is also
periodic, but the unit cell contains two junctions with different
properties, $(\omega_p,Z_J)$ and $(\omega_p',Z_J')$, which are spread
with two different spacings~[Fig.~\ref{fig:bands}c]. We find one band gap around $\omega = \omega_p$ in the first setup, and two band gaps around $\omega = \omega_p$ and $\omega = \omega_p'$ in the second, more complex case.

These one dimensional microwave photonic crystals have a variety of
applications~\cite{Joannopoulos}. The first one is the suppression of
spontaneous emission from qubits, which is achieved by tuning their
frequency to lay exactly in the middle of a band gap. Another
application is the dynamical control of group velocities. While the
width of the band gaps is more directly related to the values of $Z_J$
and the separation among scatterers, their position depends on the
scatterer frequency, $\omega_p$. Replacing the JJs with
SQUIDs~\cite{note, Haviland2000, Hutter2011}, it becomes possible to dynamically tune the
slopes of the energy bands, changing from large group velocities
(large slope) to almost flat bands (cf. Fig.~\ref{fig:bands}d) where
photons may be effectively frozen \cite{Shen2007a}. Flat bands may themselves be used to create quantum memories and also to induce a tight-binding model on the photons, in the spirit of coupled-cavity systems~\cite{hartmann06,angelakis07}. A third application is the engineering of dissipation where photonic crystals provide a new arena for theoretical and experimental studies. We will focus on this point in the last section, studying the relation between disorder, localization and entanglement generation in 1D {\it quantum circuit crystals}.

\subsection {Two-dimensional circuit crystals}
\label{sec:2D}

The evolution from one dimensional arrangements to two-dimensional or quasi-2D circuit crystals demands a careful analysis. The reason for this extra complication is that, unlike in 1D or tree configurations, phase quantization along closed paths 
introduces new contraints that prevent us from gauging away the \textit{static} phases $\varphi_{ij}^{(0)}$ and fluxes $\phi^{(0)} (x,y)$ in absence on travelling photons. More precisely, for any closed path $\mathrm{C}$ on the lattice we have
\begin{equation}
\label{fqxy}
\sum_{\rm C}  \varphi_{i,j} =  2 \pi n - \frac{2 \pi}{\Phi_0} \big ( \Phi_{\rm ext} + \Phi_{\rm ind} \big ),\quad n\in\mathbb{Z}
\end{equation}
Where $\varphi_{ij}$ are the phase differences along each branch and $\Phi_{\rm ext} + \Phi_{\rm ind}$ is the sum of external and induced fluxes enclosed by $C$. The presence of these fluxes may forbid an equilibrium condition with all phases equal to zero.

The physics of our two-dimensional crystals is intimately related to that of 2D Josephson junction arrays (JJAs), a system whose equilibrium and non-equilibrium properties have been thoroughly studied in the last twenty years \cite{Majhofer1991, Phillips1993, Dominguez1996,Mazo1996,  Newrock1999, Ciria1999, Mazo2002, Mazo2003}. In particular,  we know that JJAs constitute a physical realizations or the classical frustrated XY model, where frustration is similarly induced by the fluxes threaded through the 2D plaquettes.

A proper study of the photonic excitations must therefore begin by studying the {\it static} state on top of which they will propagate. For that we may rely on the classical nonlinear expression for the circuit's energy, built from capacitive and inductive terms, where the latter contain both the junctions and the (adimensional) mutual inductance matrix, $L_{ij,kl}$
\cite{Ciria1999}
\begin{equation}
\label{effHXY}
{\mathcal V} = - E_J \left [ \sum_{i,j} \cos (\varphi_{ij}) 
+ \frac{1}{2} \sum_{ij, kl} \sin (\varphi_{ij})  L _{ij, kl} \sin (\varphi_{kl})
\right ].
\end{equation}
The optimization of this problem is a formidable task: minimization of
(\ref{effHXY}) subjected at (\ref{fqxy}) when the induced flux are
related to the current (phases) through the inductance matrix.  In
fact, there is no known solution if any DC field is applied. However,
let us focus on a setup without external fields, then
$\varphi_{ij}^{(0)} = 0$.  This is stable against small perturbations
and against the quantum fluctuations induced both by the capacitive
terms and the travelling photons because the phases are linear in the
applied field at small fields \cite{Dominguez1996} and they enter on
second order in the scattering equations [Cf. Eq. (\ref{scjjlf})].

\begin{figure}
  \centering
  \includegraphics[width=0.55\linewidth]{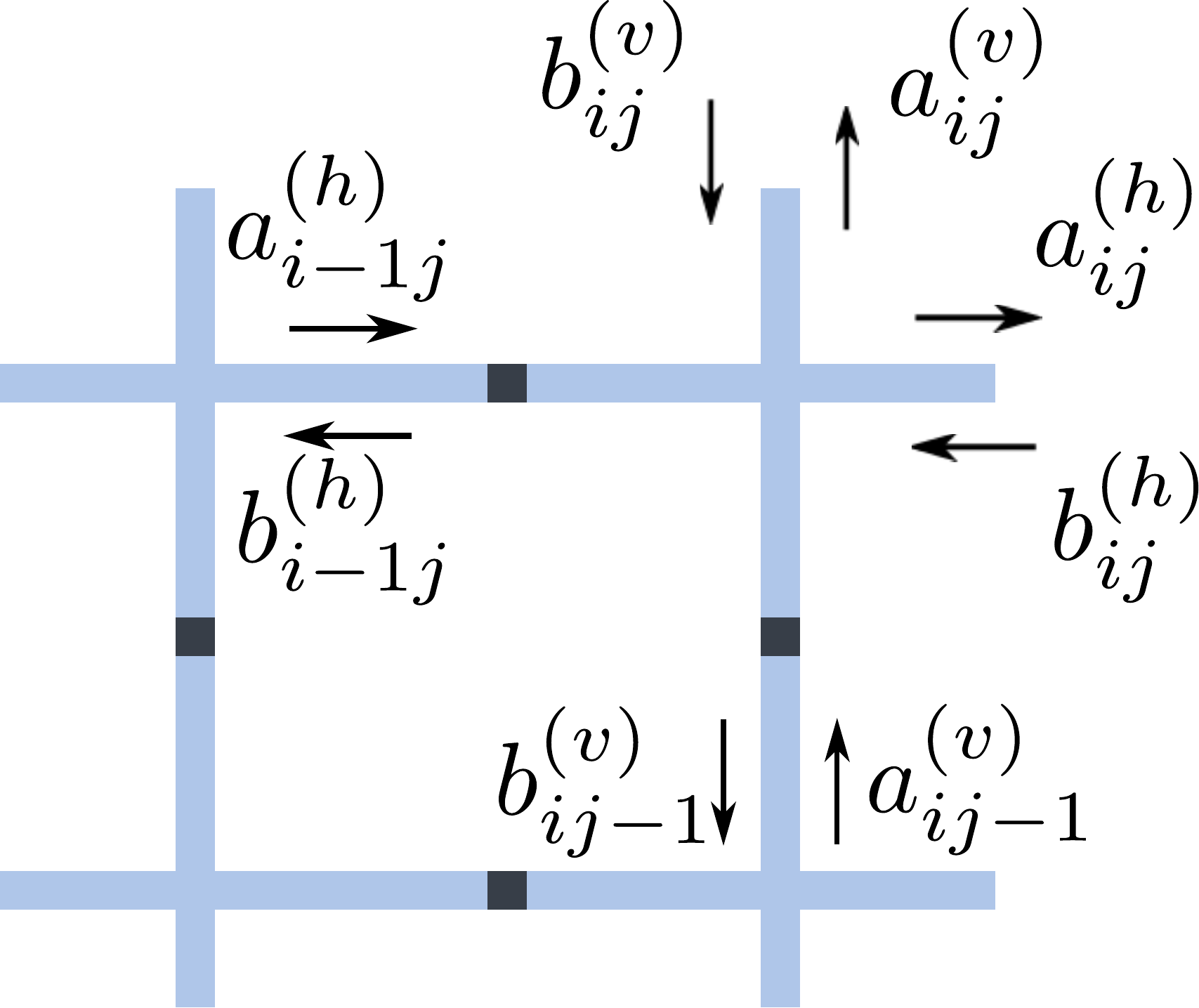}
  \caption{(color online) Scattering variables in the two-dimensional square
    lattice.  Not shown are the {\it intermediate} variables: just
    after the junctions.  In Appendix \ref{app:2D} are denoted with a bar.}
  \label{fig:sketch2d}
\end{figure}

Starting from such stable solution $\varphi_{ij}^{(0)} = \phi^{(0)}(x,y) =0$, we can redo the linear scattering theory, which now contains {\it horizontally} and {\it vertically} propagating fields [Fig.~\ref{fig:sketch2d}], 
\begin{eqnarray}
\tilde \phi_{ij} ^{(h)} &=&  a_{ij} ^{(h)} {\rm e} ^{i k_x x}  + b_{ij} ^{(h)} {\rm e} ^{-i k_x x} \label{eq:2dvars}
\\
\tilde \phi_{ij} ^{(v)} &=&  a_{ij} ^{(v)} {\rm e} ^{i k_y y}  + b_{ij} ^{(v)} {\rm e} ^{-i k_y y} \nonumber
\end{eqnarray}
with $\textbf{k} = (k_x, k_y)$.  Pretty much like in the one dimensinal case, invoking periodicity the solutions are Bloch waves,
\begin{equation}
\label{Bloch2D}
  \left(
    \begin{array}{l}
      a^{(h,v)}_{ij} \\ b^{(h,v)}_{ij}
    \end{array}
  \right) =
  \sum_{\bf p} {\rm e}^{i {\bf p}{\bf u}}
\left (
    \begin{array}{l}
      a_{\bf p}^{(h,v)} \\ b_{\bf p}^{(h,v)}
    \end{array}
  \right).
\quad
{\bf u} = (i,j)
\end{equation}
To obtain the condition for the quasimomentum ${\bf k}= {\bf p} d$ we relate the fields in ($i,j$) with the ones at ($i-1,j$) and ($i,j-1$)  as marked in figure \ref{fig:sketch2d}. Together with (\ref{Bloch2D}) we end up in a homogeneus set of linear equations, see Appendix \ref{app:2D} for the explicit calculation  .
When both the horizontal and vertical branches are equivalent,  this simplifies to [cf. Eq. (\ref{bandscond1D})]
\begin{equation}
\label{bandscond2D}
\cos (p_x) + \cos(p_y) = {\rm Tr} [ \hat T_{\rm cell} (\omega)]
\end{equation}
based on the transfer matrix along one horizontal or vertical branch,  $ \hat T_{\rm cell} (\omega)$, from each elementary plaquette.

\subsection{Two-dimensional arrays and negative index of refraction}

Once we have the possibility of building two-dimensional circuit crystals, we can also study the propagation of microwaves on extended metamaterials, or on the interface between them, with effects such as evanescent waves (i.e. localized modes) and refraction.

\begin{figure}
  \centering
  \includegraphics[width=0.9\linewidth]{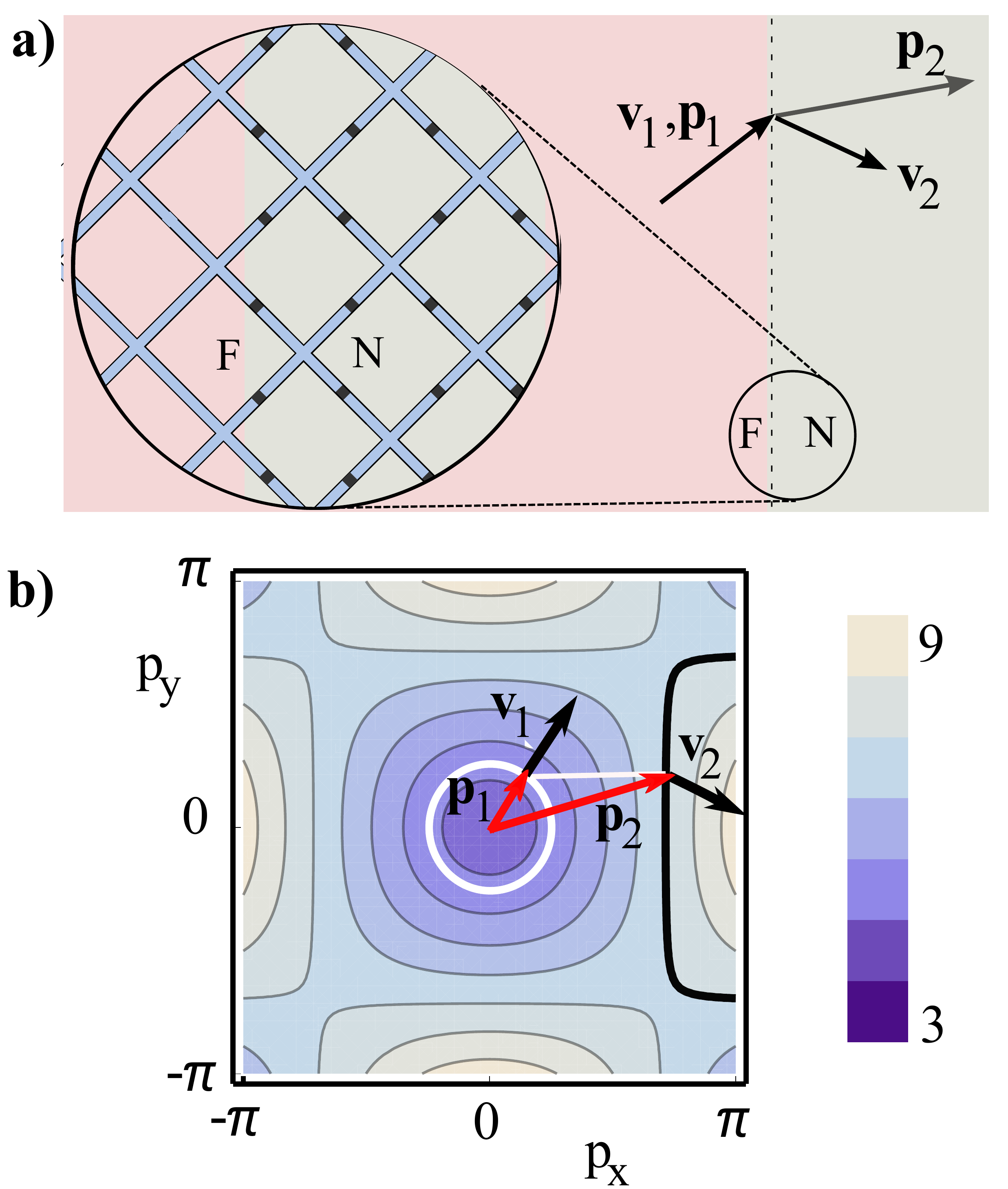}
  \caption{(color online) Two-dimensional circuit crystal. (a) Interface between a
    junction free region (F) and an engineered bandgap region (N)
    containing junctions with $\omega_J=1.1,\;Z_0/Z_J=0.8$ and
    $d=0.1$. Like in the case of polaritons in
    Ref. \onlinecite{Quach2011} the square lattice must be
    $\pi/4$-rotated to have negative refraction.  (b) Dispersion relations for the region N. In order to find out the refraction angle for wave that propagates from F to N we have to match, on each side of the interface, both the photon frequency and the projection of the wave vector along the boundary, $\mathbf{v}_1 \cdot \mathbf{e}_y = \mathbf{v}_2 \cdot \mathbf{e}_y$. For large enough momenta, (b) shows that the wave gets reversed, that is, the index of refraction is negative.}
\label{fig:2dgrid}
\end{figure}

The setup we have in mind is sketched in Fig.~\ref{fig:2dgrid}a, where we draw a two-dimensional array of lines with an interface separating a region with junctions (N) from a region where photons propagate freely (F). We may study how an incoming wave that travels against the boundary enters the N region, inducing reflections, changes of direction and attenuation. For simplicity we will assume that the free region is associated to a vacuum with linear dispersion relation $\omega_F^2 = c_1^2 (k_x^2 + k_y^2)$, where $c_1$ the effective velocity of light. The region with junctions, on the other hand, has an engineered dispersion relation $\omega_N(k_x,k_y)$, as discussed above.

When a wave hits the interface between both regions it may reflect and
refract. The wave that penetrates N has to satisfy two constraints:
the frequency of the photons must be the same as in F and the
component of the wavevector which is parallel to the interface ($p_y$
in Fig.~\ref{fig:2dgrid}a) also has to be conserved. Both constraints
arise from a trivial matching of the time ($\omega$) and spatial
($p_y$) dependence of both waves. Following
Ref.~\onlinecite{Quach2011} both constraints may be solved by
inspecting the dispersion relation in a contour plot
[Fig.~\ref{fig:2dgrid}b]. Once the matching values of the momenta are
found [red in Fig.~\ref{fig:2dgrid}] the effective group velocities
may be computed to determine the trajectory of light. As shown in that
plot, in regions where the dispersion relation is convex, the velocity
may change orientation and give rise to a refracted angle $\theta_R =
\arctan (\tan p_{1,y} \cot p_{2,y})$ whose associated index of
refraction is negative. 

The previous phenomenology has also been proposed for a related platform that consists on a two-dimensional array of coupled atom-cavity systems. Working in the single polariton subspace, it is possible to derive the dispersion relation for those artificial photons~\cite{Quach2011} and model the array as an effective photonic material. For a similar band structure and interface as the one shown in Fig.~\ref{fig:2dgrid}, the {\it effective} electrical and magnetic permitivity become negative, and one obtains again a negative reffraction angle~\cite{Veselago1968}. The engineering of these counterintuitive refraction processes is of great interest in the field of linear optics, as negative indices allow designing perfect lenses~\cite{Pendry2000}, but the propagation of photons in these mixed materials may be interesting also for engineering the dynamics of photon wavepackets, photon routing and 100\% efficient qubit-qubit interactions ---based on the perfect refocusing properties of these metamaterials.

\section{Qubit-Crystal interaction:   Quantum Master Equation approach}
\label{sec:qme}
So far we have discussed lines with junctions for tailoring photonic
transport.  In this section we study the interaction of these
metamaterials with superconducting qubits.   Modifying light-matter
interaction is a cornerstone in Quantum Optics. One of the most famous examples is the Purcell effect. Confined field enhances or dismisses the spontaneous emission for a quantum emitter.  Confinement is usually accomplished by reducing the field to one dimensional waveguides or within cavities or resonators, see {\it e.g.} Ref. \onlinecite{Cohen-Tannoudji1992}.  Related to this is the suppression of spontaneus emission when the transition frequency for the qubit is placed inside the gap of a photonic crystal \cite{John1994}.  While the first is at the heart of current circuit QED experiments, the second can be observed with our proposal for engineering band gaps.  In this section we modelize such light-matter interaction. In the weak coupling limit we work a quantum master equation and write it in terms of the Green function for the line.  The latter can be calculated by knowing the scattering matrix (\ref{Tmat}).  Finally we give a novel application, as the entanglement generation through disorder. To simplify the discussion, and without loss of generality, we focus in the one dimensional case. 

Let us write the qubit-line hamiltonian~\cite{Shen2005, hoi2011} ,
\begin{equation}
    H_{\rm tot} = H_{\rm q} + H_{\rm line} + H_{\rm int},
\end{equation}
$H_{\rm q}$ is the qubit Hamiltonian and $H_{line} = \int {\rm d}\omega \omega a^\dagger_\omega a_\omega$ the line expressed in second quantization. The qubit-photons interaction is given by
\begin{equation}
\label{Hint}
H_{\rm int} =  \hbar \sigma^x \int {\rm d} \omega g (x, \omega) (a_\omega + a_\omega^\dagger )
\end{equation}   
where $g (x, \omega)$ is the coupling per mode. In Appendix \ref{app:qlint} we show that this coupling can be expressed in terms of the Green function for the line, $G(x,y,\omega)$, as
\begin{equation}
\label{gw}
| g(x,\omega)|^2 =
2 \pi 
\,\frac{g^2}{v} \,
{\rm Im} G(x,x,\omega),
\end{equation}
where $v = 1 /\sqrt{l_0 c_0}$ is the light velocity and $g$ is the
coupling in a $\lambda/2$ superconducting resonator with fundamental
frequency $\omega$.  This is a very convenient way of expressing the
qubit-photon interaction because of two reasons. First of all,
calculation is simplified to the computation of Green functions, wich
in our case is particularly easy, as it can be derived from the the
transfer matrix (\ref{aTa}) as explained {\it e.g.} in
Ref.~\onlinecite{Tai} [See Appendix \cite{app:G} for further details] . Second and equally important, the strength of the coupling is parameterized by a simple number, $g$, which corresponds to a measurable quantity in qubit-cavity experiments ---ranging from few to hundreds MHz, from strong to ultrastrong coupling regimes.

\subsection{Qubits Quantum Master Equation}

Let us introduce a master equation for $N_q$ identical qubits placed in the line at positions $x_j$.  The qubits do not directly interact, but they will do through the line. 
 The qubit Hamiltonian reads,
\begin{equation}
\label{Hq}
H_{\rm q}  = \hbar \, \frac{\epsilon}{2} \, \sum_j^{N_q} \sigma^z_j
\end{equation}
Interested as we are here in the qubit dynamics, we can trace out the transmission line. Assuming, for simplicity that the weak coupling qubit-line limit holds,  one ends up with a master equation for the two qubit reduced density matrix~\cite{Breuer, Rivas}
\begin{equation}
\label{qme}
\frac{\partial\varrho}{\partial t}
= - \frac{i}{\hbar} [ H_{\rm q} + H_{\rm LS}, \varrho]
- \sum_{i, j=1}^{N_q} \gamma_{ij} \left ( [\sigma_i^+,\sigma_j^- \varrho ] + \mathrm{H.c.}\right ).
\end{equation}
Here, 
\begin{equation}
\label{HLS}
H_{\rm LS} =  \hbar \sum_{i  j}^{N_q} J_{ij} ( \sigma^+_i \sigma^-_j + \sigma^-_j \sigma^+_i)
\end{equation}
is the coherent coupling mediated by the line, the so called  \textit{Lamb shift} with,
\begin{equation}
\label{Jij}
 J_{ij} = \frac{g^2}{v \epsilon^2}
 {\mathcal P}
 \left [
 \int {\rm d} \nu
 \frac{\nu^2 {\rm Im} G (r_i, r_j, \nu)}{\epsilon - \nu}
 \right ]
\end{equation}
Finally the rates $\gamma_{ij}$ read,
\begin{equation}
\label{gij}
\gamma_{ij} = 2 \pi \frac{ (\hbar g)^2}{v} {\rm Im} G (r_i, r_j, \epsilon) + \lambda \delta_{ij}
\end{equation}
with $\lambda$ the  phenomenological non-radiative rate, coming from the intrinsic losses of the qubits, see Appendix \ref{app:qlint} for details.

The simplest situation that is described by this model is that of an
open transmission line, with no intermediate scatterers. In this case
the line gives rise to both both a coherent and incoherent coupling,
quantified by $J_{ij}$ and $\gamma_{ij}$ respectively, which depend on
the wavelength of the photons, the qubit separation and their
energies.  In this case without junctions, $G(x_i, x_j, \omega) = i (v/2 \omega) {\rm e}^{i \omega/v |x_i - x_j|}$ and thus
\begin{eqnarray}
\gamma_{ij} = (v/2 \omega) \cos[\omega/v (x_i - x_j)],\\
J_{ij} = (v/2 \omega) \sin[\omega/v (x_i - x_j)].\nonumber
\end{eqnarray}
Note how each of these couplings can be set independently set to zero. This has been used for modify the qubits emission going from superradiance and subradiance~\cite{Tudela2011}, and it describes recent results in multi-qubit photon scattering~\cite{vanloo12}.

\subsection{Entanglement through disorder}%
\label{sec:ed}

With all this theory at hand we move to study a concrete example where we put together structured lines, qubits, disorder and entanglement.  So far we discussed regular (periodic) arrangments of junctions producing an ideal photonic crystal. 
It is feasible to produce them, despite fabrication errors, and the
junctions within the same sample are very similar. Nevertheless it can
be interesting to induce disorder in the scattering elements, either
statically, intervening in the design or deposition processes, or
dynamically, replacing the junctions with SQUIDs and dynamically
tuning their frequencies. Disorder may have a dramatic influence in
the transport properties of the photonic
crystal~\cite{Kaliteevski2006}. On the one hand, the transmission
coefficient averaged over an ensemble of random scatterers $\langle T
\rangle$ decays exponentially with increasing length $L$ of the
disordered media, similar to Anderson's
localization~\cite{Berry1997}. On the other hand disorder  fights against the interference phenomena that gives rise to the existence of band gaps. The consequence of this competition will be that a sufficiently large disorder could restore the transmission in the frequency range that was originally forbidden~\cite{John1987, Vlasov1999, Kaliteevski2006}.  In what follows we exploit this phenomenon in connection to a purely quantum effect: the entanglement generation through disordered media.

\begin{figure}
  \centering
  \includegraphics[width=\linewidth]{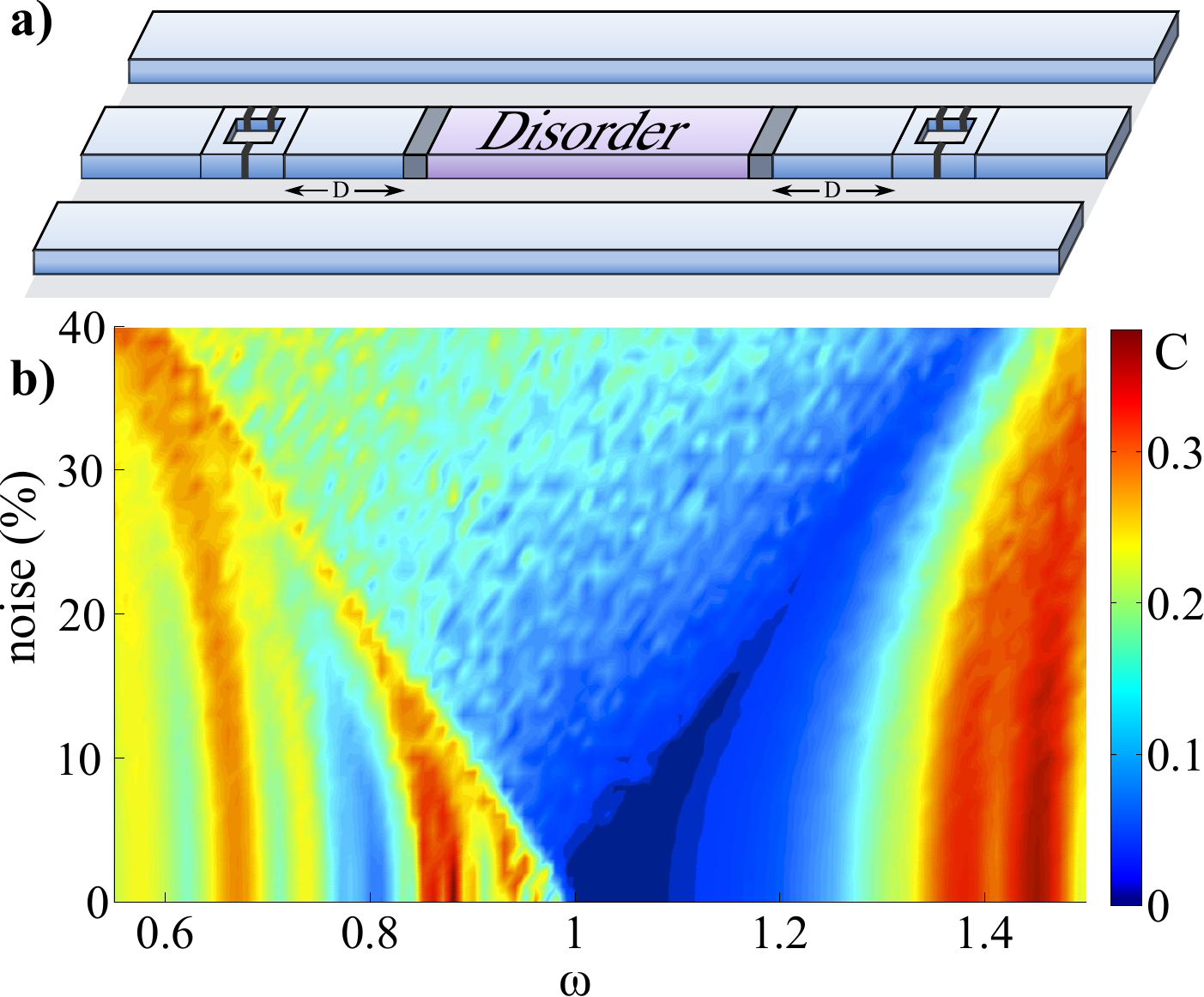}
  \caption{(color online) (a) Two qubits connected by a noisy environment. (b)
    Concurrence between the qubits for model (\ref{qme}) as a function
    of frequency and fabrication error ($\delta$). We simulated a setup with 20
    junctions regularly spaced over a distance $L=2\lambda$, averaging
    over 500 realizations. We use the parameters, $Z_0/Z_J=10$, $\epsilon
    = \omega_d$, $\lambda = 0.4 \gamma_0$ and $f = 0.1 \gamma_0$.}
\label{fig:disorder}
\end{figure}

Our model setup consists on two well separated flux qubits, $N_q=2$ in
(\ref{qme}) and (\ref{HLS}), which are
coupled by a quantum circuit disordedered media ~[Fig.~\ref{fig:disorder}]. The qubits will be at their degeneracy points and one of them is driven by an external resonant classical field:
 $\omega_{\mathrm{d}} = \epsilon$,
  \begin{equation}
  \label{Hq}
  H_{\rm q} = \frac{\epsilon}{2} (\sigma^z_1+ \sigma^z_2) + f ({\rm e}^{-i\omega_{\mathrm{d}}t} \sigma^+_1 + \mathrm{H.c.}) \; .
  \end{equation}

The line, seen now as a quantum bath,  has been interrupted by a set
of JJ forming the disordered media. The line itself follows our
previous scattering theory with uniform disorder $\delta$ in the
frequency $\omega_p \to \omega_p(1 + \delta)$ and impedance $Z_J
\to Z_J(1 + \delta)$.  We use the master equation (\ref{qme})
and compute the coefficients in the case of two qubits separated by disorder as in figure \ref{fig:disorder}.  As demonstrated
in Appendix \ref{app:G} the final expressions read,
\begin{equation}
 J_{12}=\gamma_0 {\rm Im}(T \exp(-i k 2 D))/2 \; ,
 \end{equation}
 accounting for the coherent coupling with the cross-dissipation  rate 
 \begin{equation}
 \gamma_{12}= \gamma_0{\rm Re}(T\exp(-i k 2 D)) \; .
 \end{equation}
 These interactions compete with the individual decay rates of the qubits,
 \begin{equation}
   \gamma_{ii}= \gamma_0(1 - {\rm Re} (R \exp(-i k D))) + \lambda \; ,
 \end{equation}
 which includes a phenomenological non-radiative decay channel,
 $\lambda$ coming from the intrinsic losses of the qubits [Cf. Eq. (\ref{gij})]. In all
 these formulas appear the effective rate $\gamma_0 = \pi \hbar^2
 g^2/v k$, the total transmission and reflection  $T$ and $R$ at the
 boundaries of the disordered part, and the qubit-disorder separation,
 $D$.  In figure \ref{fig:disorder} this dependence dissapears since
 the results are drawn  at the distance $D$ that maximizes the concurrence.

The physical picture that results is intuitively appealing: for the qubits to be entangled, the noisy environment should be able to transmit photons, $T\neq 0$, as both the coherent and incoherent couplings depend on it. Moreover, all photons which are not transmitted but reflected add up to the ordinary spontaneous emission rates of the qubits, $\gamma_{ii}$.  And finally, for a wide parameter range the two qubits are entangled also at $t\to\infty,$ in the stationary state of the combined system, $\partial_t \varrho_{\rm stationary} =0$. We have quantified the asymptotic amount of entanglement using the concurrence, $C$, for a variety of disorder intensities in a medium which is composed of $N=20$ junctions which are uniformly spread over a distance $L=2\lambda$. Fig.~\ref{fig:disorder} shows the result of averaging 500 realizations of disorder and contains the two ingredients stated above. We observe that for zero or little disorder entanglement becomes zero at the band gap, $\omega/\omega_p = 1$, where photons are forbidden due to interference. However, as we increase disorder the gap vanishes and entanglement enters the region around it. Outside the gap the effect is the opposite: disorder reduces the amount of entanglement, as it hinders the transmission of photons. To understand the modulations of the plot one must simply realize that the value of $C$ mostly depends on the ratio between $\gamma_{12}$ and $\gamma_{ii}$~\cite{Tudela2011}, and these are complex functions of $T$ and $R$, respectively~\cite{Chang2011a}.

\section{Conclusions and outlook}

In this work we have developed an architecture for quantum metamaterials based the scattering of travelling photons through Josephson junctions. We have shown that a single junction acts as a perfect mirror for photons that resonate with its plasma frequency. Using the scaterring matrix formalism, we have studied the band structure of networks of transmission lines with embedded junctions. We demonstrate that these setups behave as quantum metamaterials that can be used to control the propagation of individual photons. This opens the door to the usual applications of {\it classical metamaterials}, such as cloaking or subwavelength precision lenses. In particular, as an illustration of the formalism for two-dimensional networks, we discussed the observation of a negative index of refraction.

We want to remark that the utility of junction quantum metamaterials extends beyond the classical regime, with interesting applications in the fields of quantum information and quantum circuits. Replacing individual junctions with tunable SQUIDs opens the door to the dynamical control of band gaps, or the generation of flat bands, which is useful for stoping light, implementing quantum memories and what would be the equivalent of coupled cavities arrays.

Two important applications of this tunability are engineering of disorder and dissipation. In the first case the focus is on the photons that travel through the network, while in the second case the focus is on how this network acts on few-level systems that are embedded in them. We combine both approaches by developing the theory for multi-qubit interactions in a quantum metamaterial. The resulting master-equation formalism combines the effects of spontaneous emission in the artificial material, with the interaction mediated by the exchange of photons. We show that two competing effects ---Anderson localization suppresses transport, but disorder populates the band gaps with localized states---, lead to the generation of stationary entanglement in these setups.

We strongly believe that this architecture is within reach for the experimental state of the art. Building on very simple components, it offers a great potential both for quantum information with flying microwave qubits (photons), and for the static and dynamic control of stationary qubits. In the near future we wish to explore the application of this technology as a replacement for the coupled-cavity architecture, where the one- or two-dimensional network replaces the cavities, offering new possibilities of tunability and variable geometry.

Finally, we want to remark that shortly after the rewrite of this manuscript, a related work appeared that develops a similar formalism for Josephson junctions embedded in transmission lines~\cite{bourassa12}.

\acknowledgments
We acknowledge Frank Deppe, Carlos Fern\'andez-Juez and  Luis Mart\'{\i}n-Moreno for
discussions. This work was supported by Spanish Governement projects
FIS2008-01240, FIS2009-10061,  FIS2009-12773-C02-01  and FIS2011-25167
cofinanced by FEDER funds. CAM research
consortium QUITEMAD, Basque Government Grants
No. IT472-10, No. UPV/EHU UFI 11/55 and PROMISCE, SOLID and CCQED
European projects.


\begin{appendix}

\section{Transfer Matrix in the 2D case}
\label{app:2D}

We detail here the calculations needed to obtain condition (\ref{bandscond2D}) in the main text. 
The idea is to relate the {\it horizontal} and {\it vertical} fields (\ref{eq:2dvars}) on both sides of each junction, as shown in Fig.~\ref{fig:sketch2d}. 
Introducing the vectors 
\begin{equation}
\mathbf{w}^{(h,v)}_{\bf d} = 
\left (
\begin{array} {c}
a_{\bf d}^{(h,v)}
\\
b_{\bf d}^{(h,v)}
\end{array}
\right ) \; ,
\end{equation}
the fields at both sides of the junction are related, see  Eq.~(\ref{aTa}),
\begin{align}
\label{app:T}
  \bar{\mathbf{w}}_{ij}^{(h)} &=  T^{(h)}_{\rm cell}(\omega) \mathbf{w}_{i-1j}^{(h)}  ,\;  
\\
\bar{\mathbf{w}}_{ij}^{(v)} &=  T^{(v)}_{\rm cell}(\omega) \mathbf{w}_{ij-1}^{(v)}  ,\;  
\end{align}
Finally we get the final $\mathbf{w}_{ij}^{(h,v)} $ by resorting to continuity and current conservation in the corner. 
It is convenient then to define the vectors $\mathbf{e}_\pm^t = (1,\pm 1)$.  Using this notation, the continuity condition is written as follows
\begin{equation}
\mathbf{e}_+^t T^{(h)}_{\rm cell} \mathbf{w}_{i-1j}^{(h)} =
\mathbf{e}_+^t T^{(v)}_{\rm cell} \mathbf{w}_{ij-1}^{(v)} =
\mathbf{e}_+^t \mathbf{w}_{ij}^{(h)} = \mathbf{e}_+^t \mathbf{w}_{ij}^{(v)},
\end{equation}
and current conservation at the corners reads
\begin{equation}
\mathbf{e}_-^t T^{(v)}_{\rm cell} \mathbf{w}_{i-1j}^{(h)} +
\mathbf{e}_-^t T^{(h)}_{\rm cell} \mathbf{w}_{ij-1}^{(v)}  =
\mathbf{e}_-^t \mathbf{w}_{ij}^{(h)} +
\mathbf{e}_-^t \mathbf{w}_{ij}^{(v)}.
\end{equation}
writting now $\mathbf{w}_{ij}^{(h,v)}$ as Bloch waves (\ref{Bloch2D}) we end up 4-coupled homogeneus set of linear equations:
\begin{equation}
{\mathbb M} (\omega)
\left (
\begin{array}{c}
\mathbf{e}_+^t \mathbf{w}_{\bf p}^{(h)}
\\
\mathbf{e}_-^t \mathbf{w}_{\bf p}^{(h)}
\\
\mathbf{e}_+^t \mathbf{w}_{\bf p}^{(v)}
\\
\mathbf{e}_-^t \mathbf{w}_{\bf p}^{(v)}
\end{array}
\right )
= 0
\end{equation}
with,
\begin{widetext}
\begin{equation}
{\mathbb M} (\omega) = 
\left (
\begin{array}{cccc}
1 & 0  & -1 & 0
\\
1 - {\rm e}^{- i k_x} \mathbf{e}_+^t T_{\rm cell}^{(h)} \mathbf{e}_+ & - {\rm e}^{- i k_x} \mathbf{e}_+^t T_{\rm cell}^{(h)} \mathbf{e}_- & 0 & 0
\\
1 & 0 & - {\rm e}^{- i k_y} \mathbf{e}_+^t T_{\rm cell}^{(v)} \mathbf{e}_+ &  - {\rm e}^{- i k_y} \mathbf{e}_+^t T_{\rm cell}^{(v)} \mathbf{e}_-
\\
1 - {\rm e}^{- i k_x} \mathbf{e}_-^t T_{\rm cell}^{(h)} \mathbf{e}_+
&
1 - {\rm e}^{- i k_x} \mathbf{e}_-^t T_{\rm cell}^{(h)} \mathbf{e}_-
&
 {\rm e}^{- i k_y} \mathbf{e}_-^t T_{\rm cell}^{(v)} \mathbf{e}_+
 &
 1 - {\rm e}^{- i k_y} \mathbf{e}_-^t T_{\rm cell}^{(v)} \mathbf{e}_-
\end{array}
\right )
\end{equation}
together with the relations due to the scattering matrix (\ref{Tmat}) properties,
\begin{align}
(\mathbf{e}_+^t T_{\rm cell}^{(h, v)} \mathbf{e}_+) (\mathbf{e}_-^t T_{\rm cell}^{(h, v)} \mathbf{e}_-)
-
(\mathbf{e}_+^t T_{\rm cell}^{(h, v)} \mathbf{e}_-) (\mathbf{e}_-^t T_{\rm cell}^{(h, v)} \mathbf{e}_+)
&= 1
\\
(\mathbf{e}_+^t T_{\rm cell}^{(h, v)} \mathbf{e}_+) + (\mathbf{e}_-^t T_{\rm cell}^{(h, v)} \mathbf{e}_-)
=
{\rm Tr} [T_{\rm cell}^{(h, v)}] 
\end{align}
\end{widetext}
Putting alltogether we have that  ${\rm det} [ {\mathbb M}] = 0$  yields the generalized condition for the two-dimensional case.
In the simplest case of fully symmetric configuration: $ T_{\rm cell}^{(h)} =  T_{\rm cell}^{(v)} := T_{\rm cell}$ we simply have [Cf. Eq. (\ref{bandscond1D})],
\begin{equation}
\cos (p_x) + \cos (p_y) = {\rm Tr} [ T_{\rm cell} (\omega)]
\end{equation}

\section{Modelling qubit-line interaction and Master Equation}
\label{app:qlint}

In this appendix we develop the model for the qubit-line
interaction. We will focus on flux qubits for the sake of
concreteness, but the results are analogous for other qubits.  Besides
we will discuss the master equation governing the qubits dynamics.
Finally we rewrite the formulas in terms of the Green function.

For flux qubits the coupling is inductive and can be written in  circuit and/or magnetic language as,
\begin{equation}
\label{Hint0}
H_{\rm int} = M I_{\rm qubit} \times I_{\rm line} 
= \mu B \; ,
\end{equation}
here $M$ stands for the mutual inductance, $I_{\rm qubit}$ and $I_{\rm
  line}$ are the currents and $\mu$ is the magnetic
qubit dipole, while $B$ is the magnetic field generated in the cavity.

The current in the line is given by $I_{\rm line} = \frac{1}{l_0} \partial_x \phi(x)$. We will expand this field using normal modes, $u_n(x)$, following the usual quantization $\phi(x,t) = \sum u_k (x) q_k (t)$, but imposing that $u_k$ are dimensionless~\cite{Bourassa2009} and satisfy the orthonormality condition $\int c_0  u_k \; u_l {\rm d}x = C_r \delta_{kl}$ with the average capacitance $C_r := \int c_0 {\rm d}x$. Expressing the canonical operator $q_k$ in the Fock basis $q_k = (a^\dagger_k + a_k)  \sqrt{\hbar/2 \omega_n C_r}$ gives us the final expression
\begin{equation}
\label{Ilo}
I_{\rm line} = \frac{1}{l_0}
\sum
\sqrt{\frac{\hbar}{2 \omega_k C_r}}
\partial_x u_k (x)
 (a_k^\dagger + a_k ) \; .
\end{equation}
The magnetic field-current relation is given by $B_{\rm line} = \mu_0 I_{\rm line}/\pi d$, with $d$ the distance between plates in the coplanar wave-guide. The quantized magnetic dipole for the qubit can also be expressed in terms of the qubit area, $A$ and the stationary current $I_p$ as $\mu = I_p  A \sigma^x$. Putting alltogether we find the interaction Hamiltonian (\ref{Hint0})
\begin{equation}
\label{Hin2}
H_{\rm int}
=
I_p A   \frac{\mu_0}{\pi d}  \frac{1}{l_0}
\frac{\hbar}{2 C_r}
\; 
\sigma_x \; \sum  \frac{1}{\sqrt{\omega_k}} \partial_x u_k (x)
 (a_k^\dagger + a_k ) \; .
\end{equation}
We can introduce the coupling strenght per mode with frequency $\omega_0$ \cite{Lindstrom2007},
\begin{equation}
\label{g}
\hbar g = I_p  A \frac {\mu_0 }{\pi^{3/2} d} \omega_0 \sqrt{\frac{ 1}{\hbar Z_0}} 
\end{equation}
Grouping  the constants and using the expression for $g$, Eq. (\ref{g}), we  rewrite (\ref{Hin2}),
\begin{equation}
\label{Hin3}
H_{\rm int}
=
\hbar \frac{g}{\omega_0} v^{3/2} \frac{\pi}{\sqrt{2 L }} 
\; \sigma_x \;
 \sum \frac{1}{\sqrt{\omega_k}} \partial_x u_k (x)
 (a_k^\dagger + a_k )
\end{equation}
where $v= 1 / \sqrt{l_0 c_0}$ the light velocity in the line and $\omega_0$ the fundamental frequency of  a cavity with a given  $g$ and $L$ is the Lenght.  As expected the above expression is nothing but the spin boson model.


\subsection {Green Function formalism}

It turns out useful to rewrite (\ref{Hin3}) in terms of the Green Function for the line.   We begin the discussion by recalling the field wave equation.  Equivalently to layered photonic crystals, it is sufficient to work the case of homogeneous line, since the problem we are dealing with is piecewise homogeneous\cite{Novotny}. For flux qubits the coupling is through the line current [cf. Eq. (\ref{Hint0})].  Thus it is more convenient to discuss the wave equation for the mode derivatives
\begin{equation}
\frac{1}{l_0} \partial^2_x (\partial_x u_k) = - \omega_n^2 c_0
\partial_x u_k
\end{equation}
with the orthogonality condition,
\begin{equation}
\int {\rm d}x \partial_x u_k \partial_x u_{k'} = L  k^2 \delta_{k k'}.
\end{equation}
The Green function for this Sturm-Liouville problem reads
\begin{equation}
\label{eqG}
\partial_x^2 G(x,x', \omega)
+
\frac{\omega^2}{v^2} G(x,x', \omega)
=
- \delta (x-x').
\end{equation}
It is pivotal the relation [cf. Eq. (8.114) in Ref. \onlinecite{Novotny}] 
\begin{equation}
\label{Guu}
{\rm Im} G(x,x', \omega)
=
\frac{v^4}{L}
\frac{\pi}{2}
\sum_k
\frac{\partial_x u_k(x) \partial_x u_k^*(x')}{\omega_k^3} \delta (\omega - \omega_k)
\; .
\end{equation}
By rewriting (\ref{Hin3}) in the continuum limit,
\begin{equation}
\label{app:Hint}
H_{\rm int} =  \hbar \sigma^x \int {\rm d} \omega g (x, \omega) (a_\omega + a_\omega^\dagger )
\end{equation}   
with $g (x, \omega)$  [combining (\ref{Guu}), (\ref{Hin3}) and (\ref{app:Hint})]
\begin{equation}
\label{app:gw}
| g(x,\omega)|^2 =
2 \pi 
\,\frac{g^2}{v} \,
{\rm Im} G(x,x,\omega) \; ,
\end{equation}
{\it i.e.} equations (\ref{Hint}), (\ref{gw}) in the main text.

\subsection{Quantum Master Equation}
\label{app:qme}

Setting the temperature to zero (typical experiments are at the mK while frequencies are GHz) the qubit dynamics, after integrating the bosonic modes,  is given by the {\it standard} master equation in Linblad form wich assumes weak coupling between the line and the qubit~\cite{Breuer, Rivas},
\begin{align}
\partial_t \rho
=
& - \frac{i}{\hbar}
[H_{\rm qubit} + H_{\rm LS}, \rho]
\\ \nonumber
&+
\sum_{i,j}
\Gamma_{i,j}
\left (
\sigma^-_i \rho \sigma^+_{j} - \frac{1}{2}
\{ \sigma^+_{i} \sigma^-_{j} \rho \}
\right )
\\ \nonumber
& +
\lambda \sum_{i,j}\left (
\sigma^-_i \rho \sigma^+_{j} - \frac{1}{2}
\{ \sigma^+_{i} \sigma^-_{j} \rho \}
\right )
\end{align}
where $\{ , \}$ is the anticonmutator.  In the equation we have distinguished the contribution to the decay rates coming from the qubit-line coupling,  $\Gamma_{ij}$, from other noise sources affecting the qubits, denoted with a phenomenological strenght $\lambda$. The explicit expressions for the $\Gamma_{i,j}$ are {\it e.g.} \cite{Breuer, Rivas}
\begin{equation}
\label{Gjj}
\Gamma_{i, j} =  |g(\epsilon)|^2
\end{equation}
Finally, we also have to consider the \textit{Lamb Shift}
\begin{equation}
H_{\rm LS}= \sum J_{ij}\left (
\sigma_i^+ \sigma_j^- + \sigma_j^+ \sigma_i^-
\right )
\end{equation}
with
\begin{equation}
\label{app:Jij}
J_{ij}
=
\frac{1}{ 2 \pi \omega^2}
 {\mathcal P}
 \left [
 \int {\rm d} \nu
 \frac{\nu^2 {\rm Im} |g (\nu)|^2}{\epsilon - \nu}
 \right ]
\end{equation}
where $\mathcal P [ \; ]$ means principal value integral.

By defining,
\begin{equation}
\gamma_{ij} = \Gamma_{ij} + \lambda \delta_{ij}
\end{equation}
together with (\ref{app:gw}) we get coefficients (\ref{Jij}) and (\ref{gij}) in the main text.

\section{Green Function for an arrangement of scatterers}
\label{app:G}

In the following we find $G (x_i, x_{j}, \omega)$ for the problem discussed in the main text.  We show that $G(x_i, x_j, \omega)$ is written in terms of reflection and transmission coefficients, $R$ and $T$ respectively. We use this to express the decays and cross couplings, $\gamma_{ij}$ and $J_{12}$, in terms of the scattering parameters, making explicit the connection between the photonic transport in the line and the dynamics for the qubits coupled to it.
\begin{figure}[t]
\centering
\includegraphics[width=0.7\linewidth, angle=0]{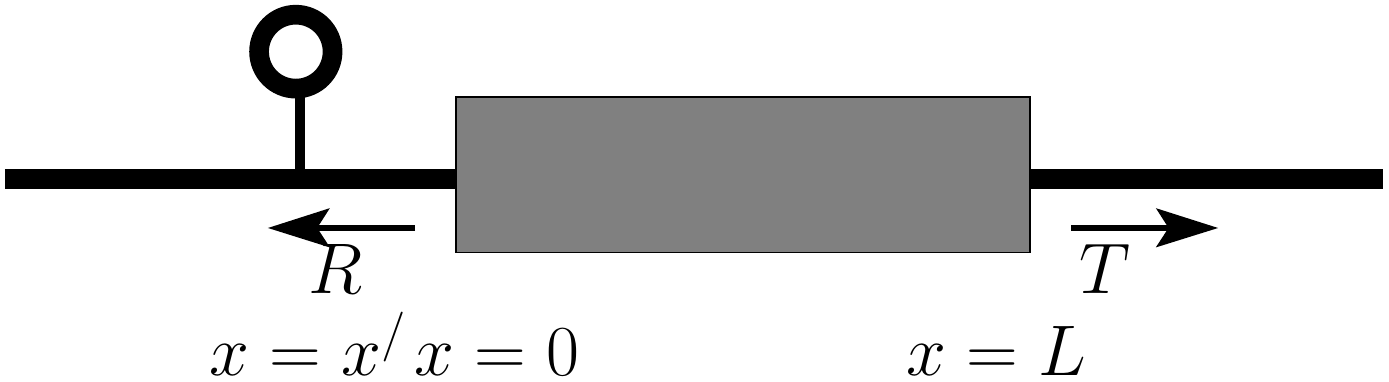}
\caption{Sketch for the Green function calculation.  The ``black box''
is characterized by transmission and reflection coefficientes.  The
source (Dirac Delta) is represented by the ring attached to the line.}
\label{fig:Green}
\end{figure}

In our case,  two qubits placed
at possitions $x_1$ and $x_2$ with a set of junctions in between (see
figure \ref{fig:Green})  $G (x_i, x_j, \omega)$ can be computed as follows.   The equation for the Green function
(\ref{eqG}) is a field equation with a source (because of the Dirac
delta) at $x=x'$. 
The junctions cover the region from  $x=0$ to $x=L$, therefore $x_1 < 0$ and
$x_2 > L$.  This situation is analogous to have a boundary with
reflection $R$ and transmission $T$, as despicted in figure \ref{fig:Green}.  In
this situation the Green Function is given by [Eqs. (2.34) and (2.35)
  in \cite{Tai}],
\begin{eqnarray}
\nonumber
G(x,x', \omega)
=
\left \{
\begin{array}{ll}
\frac{i}{2k}
\left ( 
{\rm e}^{-i k (x- x')}
-
R{\rm e}^{-i k (x + x')}
\right )
& 
x < x'
\\
\frac{i}{2k}
\left ( 
{\rm e}^{i k (x- x')}
-
R {\rm e}^{-i k (x + x')}
\right )
& 
x' < x<0
\\
\frac{i}{2k}
T {\rm e}^{i k (x- (x'+L))}
&
x > L
\end{array}
\right .
\end{eqnarray}
We remind that the minus sign in front of the $R$ above comes because the Green function in (\ref{Guu}) is given in terms of $\partial_x u_k$.

Finally, the
coefficients in the master equation read [Cf. Eqs. (\ref{gij}) and (\ref{Jij})]
\begin{align}
\nonumber
\Gamma_{jj} (\omega_{\rm qubit} ) &= 
2 \pi 
\frac{(\hbar g)^2}{v}
{\rm Im} G (r_j, r_j, \omega_{\rm qbuit})
\\
&=
2 \pi 
\frac{(\hbar g)^2}{v}
\frac{1}{2k}
\Big ( 
1 + {\rm Re} (R)
\Big )
\end{align}
\begin{align}
\nonumber
\Gamma_{12} (\omega_{\rm qubit} ) &=
2 \pi 
\frac{(\hbar g)^2}{v}
{\rm Im} G (r_1, r_2, \omega_{\rm qbuit})
\\
& =
2 \pi 
\frac{(\hbar g)^2}{v}
\frac{1}{2k}
{\rm Re} ( T  {\rm e}^{- i k (x_1 - x_2)} )
\end{align}

The last obstacle to write  the master equation is to
perform the integral in (\ref{Jij}).  Here we made use of the so
called Generalized Kramers-Kroning relation \cite{Dzsotjan2011}, namely
\begin{equation}
{\mathcal P}
\left [
\int_0^\infty
\;
{\rm d} \omega
\frac{\omega^2}{v^2}
\frac{{\rm Im} G(r_j, r_k, \omega)}{\omega - \omega_{\rm qubit}}
\right ]
=
\frac{\pi}{2} \frac{\omega_{\rm qubit}^2}{v^2}
{\rm Re} G(r_j, r_k, \omega)
\end{equation}
Thus,
\begin{equation}
J_{12}=  \pi 
\frac{(\hbar g)^2}{v}
\frac{1}{2k} {\rm Im} ( T {\rm e}^{- i k (x_1 - x_2)} )
\end{equation}

Introducing the definition,
\begin{equation}
\gamma_0 = \pi 
\frac{(\hbar g)^2}{v}
\frac{1}{k}
\end{equation}
we end up with the expressions used in the main text.




\end{appendix}


\end{document}